\documentclass[journal]{IEEEtran}
\usepackage{amssymb,amsmath,epsfig,psfrag,epstopdf,enumerate}

\usepackage[english]{babel}
\usepackage{color}
\usepackage{graphicx}
\usepackage{epstopdf}

\usepackage[normalem]{ulem}
\usepackage{varioref}
\usepackage{balance}
\usepackage{subfig}

\usepackage{fancyhdr}
\fancypagestyle{firstpage}{
  \fancyhf{}
 \fancyhead[C]{N.B. A USB memory stick, storing my draft PhD thesis - comprising material partly used in this work, "disappeared" from my hotel room safe in July 2017. For this reason, I'm unnecessarily forced to post this preprint prior publication acceptance.}
  \fancyfoot[C]{This work has been submitted to the IEEE for possible publication.  \\Copyright may be transferred without notice, after which this version may no longer be accessible.}
}

\newtheorem{definition}{Definition}[section]

\newtheorem{remark}{Remark}[section]
\newtheorem{theorem}{Theorem}[section]

\begin{document}


\title{Golden Angle Modulation}


\author{
\IEEEauthorblockN{Peter Larsson \emph{Student
Member, IEEE}}%
\thanks{The author is with the ACCESS Linnaeus Center and the School of Electrical Engineering at  KTH Royal Institute of Technology, SE-100 44 Stockholm, Sweden. E-mail: pla@kth.se.}
}
\maketitle
\thispagestyle{firstpage}

\begin{abstract}
Quadrature amplitude modulation (QAM) exhibits a shaping-loss of $\pi \mathrm{e}/6$, ($\approx 1.53$ dB) compared to the AWGN Shannon capacity.  With inspiration gained from special (leaf, flower petal, and seed) packing arrangements (spiral phyllotaxis) found among plants, a novel, shape-versatile, circular symmetric, modulation scheme, \textit{the Golden Angle Modulation (GAM)} is introduced. Disc-shaped, and complex Gaussian approximating bell-shaped, GAM-signal constellations are considered. For bell-GAM, a high-rate approximation, and a mutual information optimization formulation, are developed. Bell-GAM overcomes the asymptotic shaping-loss seen in QAM, and offers Shannon capacity approaching performance. Transmitter resource limited links, such as space probe-to-earth, and mobile-to-basestation, are cases where GAM could be particularly valuable.
\end{abstract}


\section{Introduction}
\IEEEPARstart{M}{odulation} formats, of great number and variety, have been developed and analyzed in the literature. Examples are PAM, square/rectangular-QAM, phase shift Keying (PSK), Star-QAM \cite{HanzoNgKellWebb04}, and amplitude-PSK (APSK) \cite{ThomasWeidDurr74}. Square-QAM, hereon referred to as QAM, is the de-facto-standard in existing wireless communication systems. However, at high signal-to-noise-ratio (SNR), QAM is known to asymptotically exhibit an 1.53 dB SNR-gap (a shaping-loss) to the additive white Gaussian noise (AWGN) Shannon capacity \cite{ForneyUnger98}. This is attributed to the square shape, and the uniform discrete distribution, of the QAM-signal constellation points.  Geometric and probabilistic shaping techniques have been proposed to mitigate the shaping-gap \cite{ForneyUnger98}. An early work on  geometric shaping is nonuniform-QAM in \cite{BettsCaldeLaroi94}. More recent works in this direction are, e.g., \cite{SzczecinskiAissGonzBaci06, MheichDuhaSzczMore11, XiangVale13, BuchaliIdleSchmaHu17}. Existing work on modulation schemes have, in our view, not completely solved the shaping-loss problem, nor offered a modulation format practically well-suited for the task. This leads us to examine new modulation formats.

\begin{figure}[tp!]
 \centering
 \vspace{-.3 cm}
 \includegraphics[width=9cm]{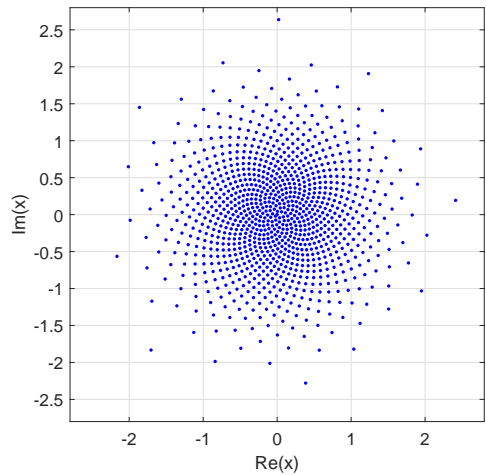}
 \caption{Geometric-bell-GAM signal constellation, $N=2^{10}$.}
 \label{fig:Fig5p5dot2}
 \vspace{0.0cm}
\end{figure}

Inspired by the beautiful, and equally captivating, cylindrical-symmetric packing of scales on a cycad cone, the spherical-symmetric packing of seeds on a thistle seed head, or the circular-symmetric packing of sunflower seeds, we have recognized that this shape-versatile spiral-phyllotaxis packing principle, found among plants, is applicable to modulation signal constellation design. Based on the spiral phyllotaxis packing, the key contribution of this letter is proposing a novel, shape-versatile, high-performance modulation framework -  the \textit{Golden angle modulation} (GAM). We consider discrete modulation, with equiprobable constellation points, that approximate a r.v. with continuous complex Gaussian (bell)-shaped distribution, as well as a baseline case, with a disc-shaped distribution. 
We find, as expected, that the MI-asymptote of geometric-bell GAM (GB-GAM), for increasing number of constellation points, coincide with the AWGN Shannon capacity. This also supports the complex Gaussian communication signal assumption, used in many performance analysis works, for the low-to-high SNR-range.

\section{Golden Angle Modulation}
\label{GAM}
The core design of GAM is given below.
\begin{definition} (Golden angle modulation)
\label{def:Def5p5d1}
The $n$th constellation point of GAM has the probability of excitation $p_n$, and the complex amplitude
\begin{align}
x_n&=r_n\mathrm{e}^{i 2\pi \varphi n}, \, n\in\{1,2,\ldots,N\},
\end{align}
where $r_n$ is the radius of constellation point $n$, $2\pi\varphi$ denotes the golden angle in rads, and $\varphi=1-(\sqrt{5}-1\thispagestyle{firstpage}
\thispagestyle{firstpage}
)/{2}$.
\end{definition}
We will assume that $r_{n+1}>r_n$ for an increasing spiral winding. For the probability, it may be equiprobable, $p_n=1/N$, or dependent on index $n$. The later, where $p_{n+1}\leq p_n$, corresponds to GAM with probabilistic shaping and is explored extensively in \cite{Larsson17}.
Hence, a constellation point, is located $\varphi\approx0.381$ turns ($137.5^{\circ}$) relative to the previous constellation point. Replacing $\varphi$, with $(1+\sqrt{5})/2\approx 1.618$, the golden ratio, or its fractional part, gives an equivalent spiral winding, but, in the opposite direction. Note that phase rotation value deviating with approximately 1\% from the golden angle (ratio) destroys the locally uniform packing.
The mathematical design of the phase rotation in Def. \ref{def:Def5p5d1} is inspired from the work by Vogel, \cite{Vogel79}, who described an idealized growth pattern for the sunflower seeds, $x_n=\sqrt{n}\mathrm{e}^{i 2 \pi \varphi n}$ (in our notation). Vogel did however not consider modulation. More importantly, a key insight here, enabling the approximation of a complex Gaussian pdf, is to not restrict the radial function $r_n$ to $\sqrt{n}$ as in \cite{Vogel79}. Allowing for arbitrary radial growth of $r_n$, gives the geometric shaping capability, and allowing for arbitrary probabilities $p_n$, gives the dimension of probabilistic-shaping.

GAM features the following advantages:
\begin{itemize}
\item
Natural constellation point indexing: In contrast to QAM, APSK, and other, without any natural index order, GAM enables a unique indexing based on signal phase, $2\pi \varphi n$, or magnitude, $r_n$, alone.
\item
Practically near-ideal circular design: A circular design can offer enhanced MI-, distance-, symbol error rate- and PAPR- performance over a square-QAM design.
\item
Shape-flexibility in radially distribution of constellation points while retaining an evenly distributed packing: We recognize this as a central feature of GAM which allows approximation of (practically) any radial-symmetric pdfs.
\item
Naturally lends itself for circular-symmetric probabilistic shaping: We recognize this as a central feature of GAM which allows approximation of (practically) any radial-symmetric pdfs.
\item
Any number of constellation points, while retaining the overall circular shape: This gives full flexibility, e.g., in alphabet size of a channel coder, or a probabilistic shaper.
\item
Rotation (and gain) invariant: The uniquely identifiable gain and rotation of signal constellation could, e.g., allow for blind channel estimation.
\end{itemize}

Some comments. First, the index range is not necessarily limited to $n\in\{1,2,\ldots,N\}$. Second, GAM has a complex valued DC component. Possible remedies, if a problem, is to subtract the DC component, or negate every second symbol. Third, while hexagonal packing is the densest 2D-packing (as desirable in the high SNR-range), it does not share the above listed features of GAM.

\subsection{Disc-GAM}
\label{sec:Sec5p5d2d1}
We first introduce disc-GAM, i.e. GAM without shaping, below. Besides its own merits, this also serves as a base-line to shaping. If the expression for $r_n$ is altered (fixed), but $p_n$ is fixed (altered), geometric (probabilistic) shaping result.

\begin{definition} (Disc-GAM)
\label{def:Def5p5d2}
Let $p_n=1/N$, $N$ be the number of constellation points, and $\bar P$ be the average power constraint. Then, the complex amplitude of the $n$th constellation point is
\begin{align}
r_n&=c_\textrm{disc}\sqrt{n}, \, n\in\{1,2,\ldots,N\},\\
c_\textrm{disc}&\triangleq \sqrt{\frac{2\bar P}{N+1}}.
\label{eq:cdisc}
\end{align}
\end{definition}
In the above, \eqref{eq:cdisc} is found from the power normalization condition,
$\bar P= \sum_{n=1}^{N} p_n
 r_n^2
 = c_\textrm{disc}^2 \sum_{n=1}^{N} \frac{1}{N}n
 =c_\textrm{disc}^2 \frac{N+1}{2}$.

\begin{figure}[tp!]
 \centering
 \vspace{-.3 cm}
 \includegraphics[width=9cm]{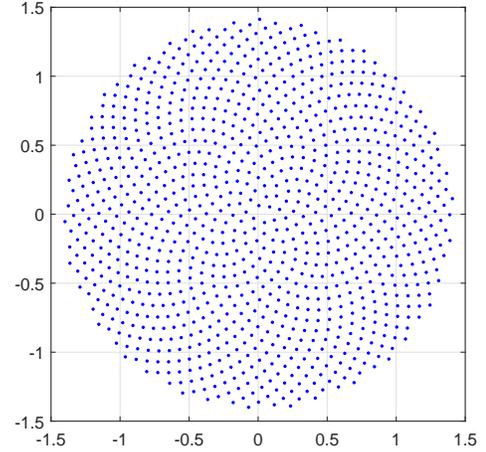}
 \caption{Disc-GAM signal constellation, $N=2^{10}$.}
 \label{fig:Fig5p5dot1}
 \vspace{0.0cm}
\end{figure}

We illustrate the disc-GAM constellation in Fig. \ref{fig:Fig5p5dot1}, where its approximate disc-shape is clearly depicted.

A few remarks about disc-GAM.
The entropy is simply $H_\textrm{disc}= \log_2 N$.
The PAPR is $PAPR_\textrm{disc}=2\bar P N /(N+1)\bar P\simeq 2$ $(\simeq 3 dB)$ when $N\rightarrow \infty$. From PAPR-point-of-view, this makes disc-GAM favorable over QAM, since $PAPR_\textrm{QAM}=4.8$ dB. (PSK has ${PAPR}_\textrm{PSK}=0$ dB, but with very poor MI-performance for $N\rightarrow \infty$).
Letting $N\rightarrow \infty$, QAM asymptotically requires $10\log_2(\pi/3)$ ($\approx 0.2$ dB) higher average power than disc-GAM for the same average constellation point distances .
Also when $N\rightarrow \infty$, QAM asymptotically requires $10\log_2(\pi/2)$ ($\approx 1.96$ dB) higher peak power than disc-GAM for the same average distance between (uniformly packed) constellation points.

\subsection{Geometric bell-GAM}
\label{sec:Sec5p5d2d2}
Next, two geometric shaped GAM schemes are introduced.

\subsubsection{High-rate approach}
\label{sec:Sec5p5d2d2d1}
This first GB-GAM design builds on the inverse sampling method (a high-rate (HR) approximation), and approximates a complex Gaussian distributed r.v.
\begin{theorem} (Geometric-bell-GAM (HR))
\label{thm:Thm5p5d1}
Let $p_n=1/N$, $N$ be the number of constellation points, and $\bar P$ be the average power constraint. Then, the complex amplitude of the $n$th constellation point is
\begin{align}
r_n&=c_\textrm{gb}\sqrt{\ln{\left(\frac{N}{N-n}\right)}},\, n\in\{0,1,\ldots,N-1\},\notag \\
c_\textrm{gb}&\triangleq\sqrt{\frac{N \bar P}{N\ln{N}-\ln(N!)}}.
\label{eq:EqGaussrn}
\end{align}
\end{theorem}
\begin{IEEEproof}
The proof is given in Appendix \ref{app:App5p5dA2d1}.
\end{IEEEproof}

We illustrate the geometric-bell-GAM signal constellation in Fig. \ref{fig:Fig5p5dot2}, and note that it is densest at its center, i.e. where the pdf for the complex Gaussian r.v. peaks.

We note the following characteristics.
When $N \rightarrow \infty$, since $\lim_{N\rightarrow \infty} {N}^{-1} \ln{\left({N^N}/{N!}\right)}=1$, we get $r_n\simeq \sqrt{\bar P \ln{\left({N}/{(N-n)}\right)}}$.
The entropy is $H_\textrm{gb}= \log_2 N$.
The PAPR is $PAPR_\textrm{gb}=c_\textrm{gb}^2\ln(N)
=\bar P /(1-\ln{(N!)}/N \ln {N})\simeq \bar P \ln{(N)}$, which tends to infinity with $N$. This is expected as the PAPR of a complex Gaussian r.v. is infinite.
Note here that the index range in \eqref{eq:EqGaussrn} can not be $n\in\{1,2,\ldots,N\}$, but is chosen as $n\in\{0,1,\ldots,N-1\}$, as $r_N=\infty$ otherwise.

\subsubsection{MI-optimization of GB-GAM: Formulation-G1}
\label{sec:Sec5p5d2d3d1}
In this method, we let $p_n=1/N$, and vary $r_n$ in order to maximize the MI, for a desired SNR $S$. The optimized signal constellation points are $x_n^*=r_n^* \mathrm{e}^{i 2\pi \varphi n}$. More formally, allowing for a complex valued output r.v. $Y$, and a complex valued (discrete modulation) input r.v. $X$, the optimization problem is
\begin{equation}
\begin{aligned}
& \underset{r_n}{\text{maximize}}
& & I(Y;X), \\
& \text{subject to}
& & r_{n+1} \geq r_{n}, \; n = \{1,2, \ldots, N\},\\
& & & r_1 \geq 0,\\
& & & \sum_{n=1}^{N}\frac{p_nr_n^2}{\sigma^2}=S.
\end{aligned}
\end{equation}

\begin{remark}
\label{rm:Rm5p5d9}
For some applications, the PAPR is of interest. A PAPR-inequality constraint, $r_N^2/\sum_{n=1}^{N-1}p_n r_n^2\leq PAPR_0$,  $PAPR_0$ being the target PAPR, can be amended to the optimization problem. Other constraints may also be of interest.
\end{remark}

When optimizing GAM in AWGN, the MI is $I(Y;X)=h(Y)-h(W)$, where $h(W)=\log_2(\pi \mathrm{e}\sigma^2)$, $h(Y)=-\int_{\mathbb{C}} f_Y \log_2(f_Y) \, \mathrm{d}y$, integrating over the complex domain,
with $f_Y=\sum_{n=1}^{N} p_n f(y|x_n)=\frac{1}{\pi\sigma^2}  \sum_{n=1}^{N} p_n \mathrm{e}^{-\frac{|y-x_n|^2}{\sigma^2}}$.

Due to the non-linearities in the MI, the optimization problem in formulation-G1 is hard to solve analytically. Hence, a numerical optimization solver is used in Section \ref{sec:Results}.

\section{Numerical Results}
\label{sec:Results}
Here, we now examine the MI-performance of GAM (with its irregular cell-shapes and varying cell-sizes) by using a Monte-Carlo simulation approach. For the optimization in formulation-G1, MATLAB's \textit{fmincon} function is used together with numerical integration for the MI.

\begin{figure}[tp!]
 \centering
 \vspace{-.3 cm}
 \includegraphics[width=9cm]{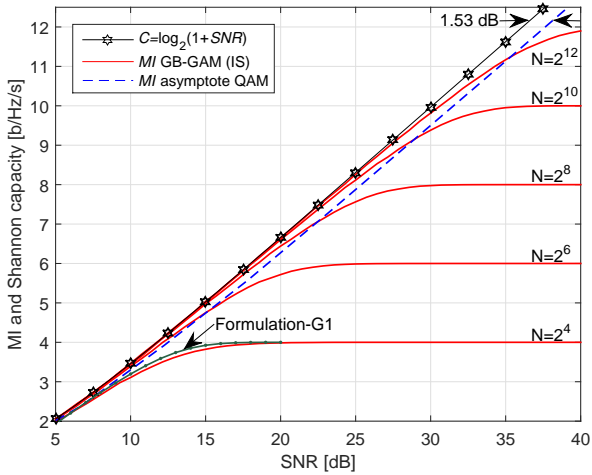}
 \caption{MI of GB-GAM (HR), asymptote of QAM, and Shannon capacity. MI of GB-GAM (G1) with $N=2^4$.}
 \label{fig:Fig5p5dot4}
 \vspace{0.0cm}
\end{figure}

In Fig. \ref{fig:Fig5p5dot4}, we illustrate the MI performance for GB-GAM (HR) together with the Shannon capacity. As expected, for larger constellation size $N$, a greater overlap with the Shannon capacity is seen. The MI approximation is good up to about $\approx2H/3$. Naturally, the MI is limited by the entropy of the signal constellation. We observe an intermediate SNR region, a region where the MI does neither reach the channel capacity, nor the entropy of the signal constellation. It is in this SNR-region that further constellation optimization, i.e. formulation-G1 , is of interest. The MI for GB-GAM (G1) is also shown, but due to optimization complexity only, for $N=16$. 

In Tab. \ref{tab:Tab1d5p5d1}, the MI of GB-GAM with the HR-, and G1-, formulations are given. As expected, the optimized scheme, G1,  perform better than HR. While the MI-improvements are modest, the optimization formulation is substantiated. For G1, when the MI is as large as the constellation entropy, we have observed that the signal constellation approaches the disc-GAM solution, whereas in the low-MI region, we have noted that the optimized signal constellation approaches the HR GB-GAM solution. It is further observed that the extreme magnitudes of the highest constellation indices for GB-GAM (HR), as seen in Fig.  \label{fig:Fig5p5dot2}, are attenuated with the MI-optimization and leads to improved PAPR performance.

\begin{figure}[tp!]
 \centering
 \vspace{-.3 cm}
 \includegraphics[width=9cm]{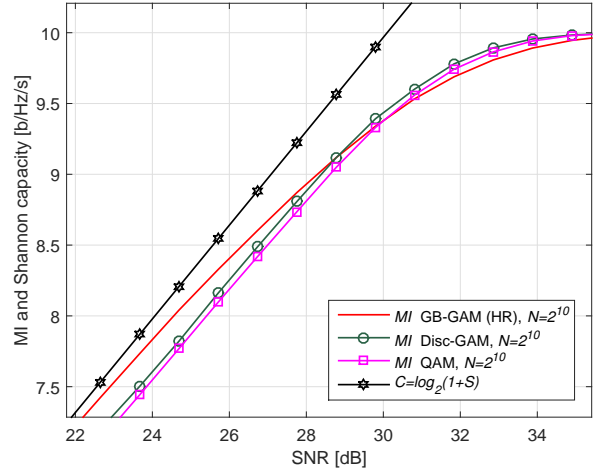}
 \caption{MI of disc-GAM, GB-GAM (HR), QAM, with $N=2^{10}$, and Shannon capacity.}
 \label{fig:Fig5p5dot6}
 \vspace{0.0cm}
\end{figure}

In Fig. \ref{fig:Fig5p5dot6}, we compare disc-GAM, GB-GAM (HR), and QAM, together with the Shannon capacity when the MI is nearly as large as the constellation entropy. We note the (expected) $\approx1.53$ dB SNR-gap for QAM to the Shannon-capacity. The (expected) SNR-gap from QAM to disc-GAM is $\approx0.2$ dB. When the MI is approximately as large as the constellation entropy, disc-GAM, and QAM, perform slightly better than GB-GAM (HR) scheme. This observation prompted us to develop the MI-optimization formulation-G1.

\begin{table}[tp!]
\normalsize
\begin{center}
  \begin{tabular}{|r *{3}{|c}|}
    \hline
        \textbf{SNR}
    &
        \textbf{AWGN capacity}
    &
        \textbf{HR}
    &
        \textbf{G1}
    \\
    \hline
    $\approx4.8$ dB&2&1.921&1.961\\
    \hline
    $\approx11.8$ dB&4&3.440&3.549\\
    \hline
    $15$ dB&$\approx 5.03$&3.828&3.926\\
    \hline
  \end{tabular}
\caption{MI in [b/Hz/s] for HR- and G1-schemes with $N=16$.}
\label{tab:Tab1d5p5d1}
\end{center}
\vspace{-0.35cm}
\end{table}

\section{Summary and Conclusions} \label{Summary}
In this letter, we have introduced a new modulation format, the \textit{golden angle modulation}.  With geometric- (or probabilistic-) shaping, GAM can approximate virtually any circular-symmetric pdf. We studied geometrically-shaped GAM to approximate the pdf of a continuous complex Gaussian r.v. A high-rate solution, was developed. We also introduced the notion of MI-optimized GAM under an average SNR-constraint, and optionally also a PAPR-inequality-constraint. The  MI-performance was observed to asymptotically approach the Shannon capacity as the number of signal constellation points tended to infinity. In contrast to QAM's 1.53 dB shaping-loss, GAM exhibit no asymptotic loss. The complex Gaussian communication signal model assumption, as often used for performance analysis, was substantiated from a practical modulation point-of-view.

We believe that GAM may find many applications in transmitter-resource-limited links, such as space probe-to-earth, satellite-to-earth, or mobile-to-basestation communication. This is so since high data-rate is desirable from the power-, energy-, and complexity-limited transmitter side, but higher decoding complexity is acceptable at the receiver side. Certain cases may also benefit from using the PAPR-inequality constrained optimization. With bell-GAM, the power reduction in the high-MI regime could be up to 30\% ($1-1/10^{1.53/10}\approx0.3$), which is of environmental interest. Moreover, cellular-system operators, could potentially also reduce energy consumption (and cost) with up to 30\%.  It is hoped that, GAM, with its attractive characteristics and performance, could be of interest for most, wireless, optical, and wired, communication systems.

\appendix 

\subsection{Proof in Theorem  \ref{thm:Thm5p5d1}}
\label{app:App5p5dA2d1}
\begin{IEEEproof}
The distribution in phase, for the constellation points, is already given by the $\mathbf{e}^{2\pi \varphi n}$-factor. However, the radial distribution need to be determined. A complex Gaussian r.v. with variance $\sigma^2$ is initially considered. The inverse sampling method, assumes a uniform continuous pdf on $(0,1)$. We modify the inverse sampling method and use a discrete uniform pmf at steps $n/N, \, n\in\{0, 1,\ldots,N-1\}$, where $N$ is assumed large (for the high-rate approximation).
\begin{align}
F(r_{n})
&=\int_0^{r_{n}}f_R(r)dr=\int_0^{r_{n}}\frac{\mathrm{e}^{-\frac{r^2}{\sigma^2}}}{\pi\sigma^2}2 \pi r \, \mathrm{d}r\notag\\
&=1- \mathrm{e}^{-\frac{r_n^2}{\sigma^2}}\notag.
\end{align}
Setting $F(r_n)=n/N$, and solving for $r_n$, yields
\begin{align}
r_n=\sigma\sqrt{\ln\left(\frac{N}{N-n}\right)}\notag .
\end{align}
Thus, the general solution for the signal constellation has the form $r_n=c_\textrm{gb} \sqrt{\ln{\left(\frac{N}{N-n}\right)}}$. The constant $c_\textrm{gb}$ is given by the average power constraint as follows,
\begin{align}
\bar P&=\sum_{n=0}^{N-1} p_n r_n^2=\frac{c_\textrm{gb}^2}{N}\sum_{n=0}^{N-1}\ln{\left(\frac{N}{N-n}\right)},\notag\\
\Rightarrow c_\textrm{gb}&=\sqrt{\frac{N \bar P}{\sum_{n=0}^{N-1}\ln{\left(\frac{N}{N-n}\right)}}}
=\sqrt{\frac{N \bar P}{N\ln{N}-\ln{N!}}}\notag.
\end{align}
\end{IEEEproof}

\end{document}